\documentstyle[nassau_preprint,pslatex,graphicx]{article}  
\frompage{21} \topage{31}                                                
\def\rightmark{Centrality Dependence of Inclusive Identified Hadrons in PHENIX}
\def\leftmark{J.M. Burward-Hoy}
\title{Centrality Dependence of Inclusive Identified Hadrons in PHENIX}
\authors{{J.M. Burward-Hoy$^{\,a}$
\index{Burward-Hoy, J.M.}
}\\[2.812mm]
\normalsize{
Physics and Advanced Technologies Directorate\\
Lawrence Livermore National Laboratory\\ 
7000 East Avenue, Livermore, California, USA
}
}

\abstract{Results of identified charged hadrons produced in Au+Au
  collisions at ${\sqrt{s_{NN}}} = 130$ GeV as a function of
  centrality are discussed.  The $\langle {p_{T}}\rangle$ of
  protons and antiprotons is significantly larger than in p--p
  collisions at the same center of mass energy. Its scaling with the
  number of participant nucleons, an increase of $20 \pm 5\%$ from
  peripheral to central, may be due to radial expansion.  In a
  hydrodynamical picture a simultaneous fit of pions, kaons and
  protons is consistent with a radial flow velocity of
  $\beta_{\,\rm{Tsurface}} = 0.72 \pm 0.01$ and a temperature of $T =
  122 \pm 4$ MeV for the 5\% most central collisions.  Dominant proton
  yields at ${p_{T}} > 1.5$ GeV/c and antiproton yields at
  ${p_{T}} > 1.8$ GeV/c may be explained by a combination of
  suppression of pion yields at high ${p_{T}}$ and the boost of low
  ${p_{T}}$ protons to high ${p_{T}}$ from radial flow.}
  

\makeindex
\begin{document}

\maketitle
\setcounter{page}{21}

\vspace*{-6pt}
\section{Introduction}\label{intro}

It is predicted by theoretical calculations that highly excited and
dense hadronic matter undergoes a phase transition from a hadronic gas
to a deconfined state of quarks and gluons called a Quark Gluon Plasma
(QGP) \cite{QGP}.  A few $\mu$s after the Big Bang \cite{BB}, before
nuclei and atoms were formed, the very hot and baryon free early
universe may have existed in such a state of deconfinement.

The objective of relativistic heavy-ion physics is to produce an
energy density in the overlap region between two colliding ions that
is high enough to probe the phase transition between hadronic matter
and quark matter.  The quarks and gluons coexisting in a QGP state
cannot be measured directly.  The experimental challenge is to study a
suite of observables that are indirect probes of plasma formation.
The one observable that is measured and discussed here are the
transverse momentum (${p_{T}}$) spectra of produced hadrons, namely:
pions, kaons and (anti)protons.  Unlike photons and dileptons, the
hadrons decouple late in the reaction due to strong interactions;
therefore, the momentum spectra are sensitive to the system at
freezeout.

If the collision can be described as an expanding, relativistic fluid,
then hydrodynamic models with various equations of state can be used
to predict the hadron spectra \cite{EOS}.  If both hydrodynamic
behavior and longitudinal boost invariance are assumed, then the
amount of expansion generated in the collision can be extracted from
the data at low momenta (${p_T} < 1.0$ GeV/c) using a
hydrodynamics-inspired parameterization.  Previously, the expansion of
matter in Pb+Pb collisions at CERN SPS energies was similarly obtained
from hadron spectra and HBT data \cite{NA44_78,NA49}.  In addition,
the measured yield of high ${p_T}$ hadrons may reveal the amount of
partonic energy loss when compared to calculations in perturbative
Quantum Chromodynamics (pQCD) parton models~\cite{hijing,PPG003}.

The hadrons from Au+Au collisions at a roughly ten times higher energy
density are measured at the Relativistic Heavy Ion Collider (RHIC) at
Brookhaven National Laboratory (BNL).  The PHENIX experiment at RHIC
is described briefly in Section~\ref{expt}.

In the first year of data taking, the identified charged hadrons
produced in five different event centralities at the collision energy
of 130 GeV in Au+Au collisions are measured:\break 0--5\%, 5--15\%,
15--30\%, 30--60\% and 60--92\% of the total inelastic cross section
of 6.8~b.  The data reduction, corrections, extrapolations, and
systematic uncertainties are summarized in Section~\ref{ana}.  The
centrality selected spectra, yields, mean transverse momenta ($\langle
{p_T} \rangle$), and total charged multiplicity are presented in
Section~\ref{results}.  An interpretation of the data based on radial,
hydrodynamic expansion is provided in Section~\ref{hydrofits}.  The
implications of the results are discussed in Section~\ref{concl}.
 
\section{Experiment}\label{expt}

\subsection{The PHENIX detector}  

\begin{figure} 
\vspace*{6pt}
\centerline{
\includegraphics[height=6.75cm]{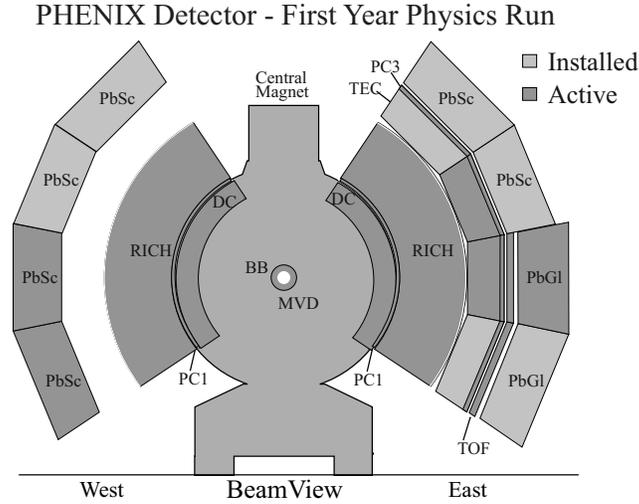}
}
\vspace*{0pt}
\caption{A cross-sectional view of the PHENIX detector, transverse 
  to the beam-line.  Within the two central arm spectrometers the
  detectors that were instrumented and operational during the
  $\sqrt{{s_{NN}}} = 130$ GeV run are shown.}
\label{fig1}
\end{figure}

The PHENIX \cite{ref1,ref2,ref3} detector is designed for the
measurement of e, $\mu$, $\gamma$, and identified hadrons over a large
momentum range for ${p_T}>0.2$ GeV/c. It is comprised of four
spectrometer arms: two at mid-rapidity called the central arms
(``east'' and ``west'') and two at more forward rapidities called the
muon arms (``north'' and ``south'').  The central arm acceptance is
$\vert \eta \vert < 0.35$ with each arm subtending $\pi/2$ in
azimuthal angle ($\phi$).  In the first year of physics running in
2000 (Run-1), half of the central arm detectors were instrumented, as
shown in Fig.\ \ref{fig1}.

The two types of collision axis detectors, the Zero Degree
Calorimeters (ZDC) and the Beam--Beam Counters (BBC) measure both the
initial time and vertex position of each event.  The ZDC resolution
was $\sigma_{{E}}/{E} < 20$\% at the single neutron energy of ${E_{n}}
= 100$~GeV with an acceptance of $|\eta|>6$.  The BBC resolution was
$\sigma_{{z}} = 1.5$ cm along the beam axis and $\sigma_{{t}} = 70$ ps
in timing, with an acceptance of $3.0<\eta<3.9$.

The time-of-flight detector (TOF) in the east arm subtends $\pi/4$ in
$\phi$ and was used for the identification of $\pi^{\pm}$, K$^{\pm}$,
and (anti)protons.  The TOF measures the arrival time of charged
hadrons at the TOF wall positioned 510 cm from the collision vertex.
In Run-1, a timing resolution of 115 ps was obtained.

In Run-1, PHENIX had two layers of multi-wire proportional chambers
with pad-readout (PC1, PC3) that measure a three-dimensional space
point of a charged track.  The points from PC1 are used in the global
track reconstruction to determine the polar angle of the track.  The
PC3 positioned in the east arm is used to minimize the background
contribution of albedo and non-vertex decay particles; it is used in
the high-${p_{T}}$ charged hadron analysis~\cite{PPG003}.

The drift chambers measure the track curvature between 202 and 246 cm
radially, for which the particle momentum is determined.  In Run-1, a
resolution of $\sigma_{{p}}/{p} = 0.6\% \oplus 3.6\%$ was obtained.
The absolute momentum scale was known better than~2\%.

\section{Data Reduction and Analysis}\label{ana}

\subsection{Event sample}

A total of 5M minimum bias triggers were recorded at
$\sqrt{{s_{NN}}}= 130$ GeV in the PHENIX ZDC detectors.  The event
centrality was determined using a correlation measurement between the
neutral energy deposited in the ZDCs and fast charged particles
($\beta \ge 0.69$) recorded in the BBCs \cite{PPG001}.  In order to
minimize the amount of albedo from the central magnet pole faces,
positioned at $\pm40$ cm, only those events with a vertex distribution
within $\pm 30$ cm were selected.  Additional event criteria include
the magnetic field setting and detector stability.  The resulting
event sample, $140\,000$ minimum bias events, was $92 \pm 4$\% of the
total inelastic cross section of 6.8~b \cite{PPG001}.  The events were
subdivided into five centrality classes: 0--5\%, 5--15\%, 15--30\%,
30--60\% and 60--92\% of the total geometric cross section.  In each
centrality selection, the number of participating nucleons
(${N_{\rm part}}$) and number of binary nucleon--nucleon collisions
(${N_{\rm coll}}$) were determined using a Monte Carlo Glauber model
calculation \cite{GLAUBER}.

\subsection{Particle identification}

Particle identification for charged hadrons was performed by combining 
the information from the tracking detectors with the timing information 
from the BBC and the TOF.  Tracks at 1 GeV/c in momentum point 
to the TOF with a projected resolution $\sigma_{\rm{proj}}$ of 5 mrad 
in azimuthal angle and 2 cm along the beam axis.  Tracks that point 
to the TOF with less than 2.0 $\sigma_{\rm{proj}}$ were selected.  

The measured momentum ($p$), path length ($L$), and time of flight
($t$) in the spectrometer were used to calculate the particle mass,
with $\beta = {L/ct}$ in Eq.\ (\ref{m2}).
\begin{equation}
{m}^2 = {p}^2 \left[ {{\left(\frac{1}{\beta}\right)}^2 - 1} \right]\,.
\label{m2}
\end{equation}
The width of the mass-squared distribution $\sigma_{{m}^{2}}$ was
calculated as a function of momentum, using the known detector
resolutions and the relation in Eq.\ (\ref{m2width}),
\begin{equation}
\sigma^2_{{m}^2}={C_1}^2\cdot 4\, {m}^4\left(1+\frac{{m}^2_0}{{p}^2}\right) +
{C_2}^2\cdot 4\frac{{m}^4_0}{{p}^2}+ 
{C_3}^2 \cdot [4{p}^2({m}_0^2+{p}^2)]\,.
\label{m2width}
\end{equation}
The constants are related to the measured drift chamber momentum
resolution (${C_1}$, ${C_2}$) and combined flight path length and
flight time errors (${C_3}$).  Tracks that fell within
2$\sigma_{{m}^{2}}$ of a measured mass centroid were identified.

\subsection{Corrections}

The raw spectra include inefficiencies from detector acceptance,
resolution, particle decays in flight and track reconstruction.
Corrections for these effects were determined using a Monte Carlo
simulation of the detector response. The baseline efficiencies were
determined by simulating and reconstructing single hadrons.
Multiplicity dependent effects were evaluated by embedding simulated
single hadrons into real events, and measuring the degradation of the
reconstruction efficiency.  For each hadron, sufficient statistics
were generated flat in azimuthal angle and rapidity for $\vert y \vert
< 0.5$ for every measured ${p_T}$-bin such that the statistical
error in the data dominates.  The statistical errors from the
correction factors were added in quadrature to the statistical errors
in the data.  Corrections for feed-down from weak decays were not
applied, but an upper limit is estimated based on a MC simulation of
$\Lambda$ particles within the PHENIX acceptance. More details on the
corrections are reported elsewhere~\cite{PPG006,PPG009}.

\subsection{Extrapolations}

The yield integrated over transverse momentum, $dN/dy$, and the
integrated average transverse momentum, $\langle {p_{T}} \rangle$,
were determined for each particle after fitting and integrating
functions that describe the spectral shape.  A power-law and
${m_{T}}$ exponential were fit to the pion data, while ${p_{T}}$
and ${m_{T}}$ exponentials were fit to each kaon and proton
spectrum.

\subsection{Systematic uncertainties}

The overall uncertainty on $dN/dy$ is 13$\%$, 15$\%$ and 14$\%$ for
pions, kaons and protons, respectively.  Uncertainties on $\langle
{p_{T}} \rangle$ depend on the extrapolation and background
uncertainties; the uncertainties are 7$\%$, 10$\%$ and 8$\%$ for
pions, kaons and protons, respectively.  An estimated 24\% of the
measured proton and anti-proton yields are due to $\Lambda$ weak
decays, based on a Monte Carlo simulation.

\section{Results}\label{results}

\subsection{Transverse momentum distributions}

\begin{figure}[ht] 
\vspace*{0pt}
\begin{minipage}[t]{8cm}
\hspace*{-3pt}\includegraphics[height=8.5cm,clip]{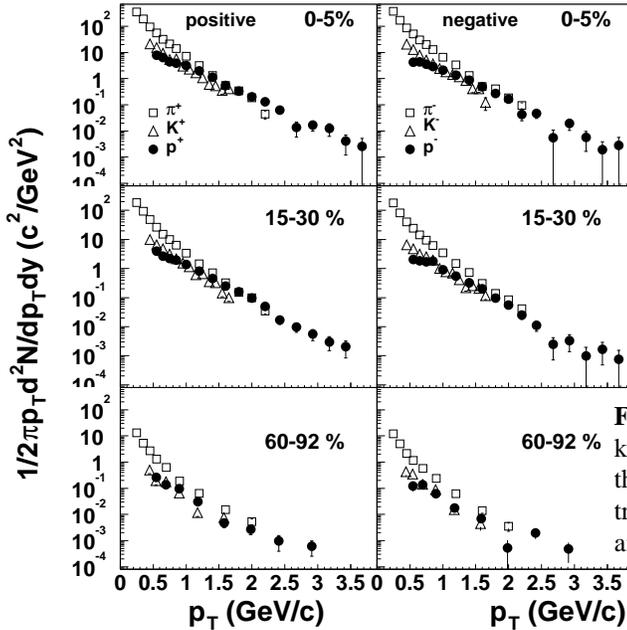}
\end{minipage}
\vfill\hspace*{8.cm}
\begin{minipage}[t]{5.25cm}
\vspace*{-109pt}
\caption{The invariant yields for pions, kaons, and (anti)protons 
  produced in three different classes of event centrality: 0--5\%
  (top), 15--30\% (middle) and 60--92\% (bottom)
  \protect\cite{PPG006}.}
\label{fig2}
\end{minipage}
\end{figure}

The $\pi^{\pm}$, K$^{\pm}$, p and $\overline{\rm p}$ invariant yields as
a function of transverse momentum (${p_{T}}$) for the most central,
mid-central, and the most peripheral events are shown in Fig.\
\ref{fig2}.  These data were reported previously \cite{PPG006}.

\begin{figure}[htb] 
\vspace*{6pt}
\centerline{
\includegraphics[width=11cm,clip]{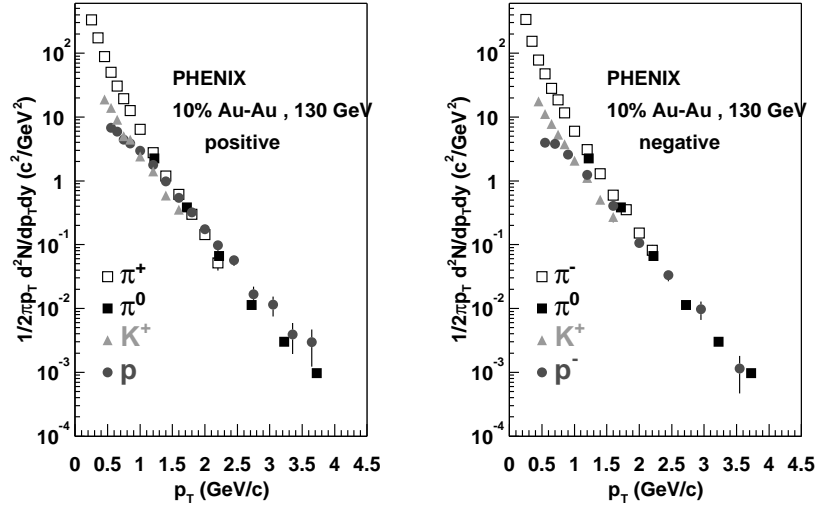}
}
\vspace*{-6pt}
\caption{The invariant yields as a function of ${p_{T}}$ for the 
  positive (left) and negative (right) pions in the most central
  events are consistent with the $\pi^{0}$ measurement from an
  independent PHENIX analysis, verifying the apparent dominance of
  baryon yields at high~${p_{T}}$.}
\label{piCompare}
\end{figure}

Transverse momentum spectra of non-identified charged hadrons and
neutral pions were reported by PHENIX in the range $1 < {p_T} < 5$
GeV/c \cite{PPG003}.  In Fig.\ \ref{piCompare}, the $\pi^{0}$ data are
shown with the charged pion data for the most central events.  The
comparison confirms the apparent baryon to meson dominance at
${p_{T}} \approx 1.5$ GeV/c and ${p_{T}} \approx 1.8$ GeV/c for
positive and negative particles, respectively.

\subsection{Yield and $\langle p_T \rangle$}

\begin{figure}[!h] 
\vspace*{6pt}
\centerline{
\includegraphics[width=11cm,clip]{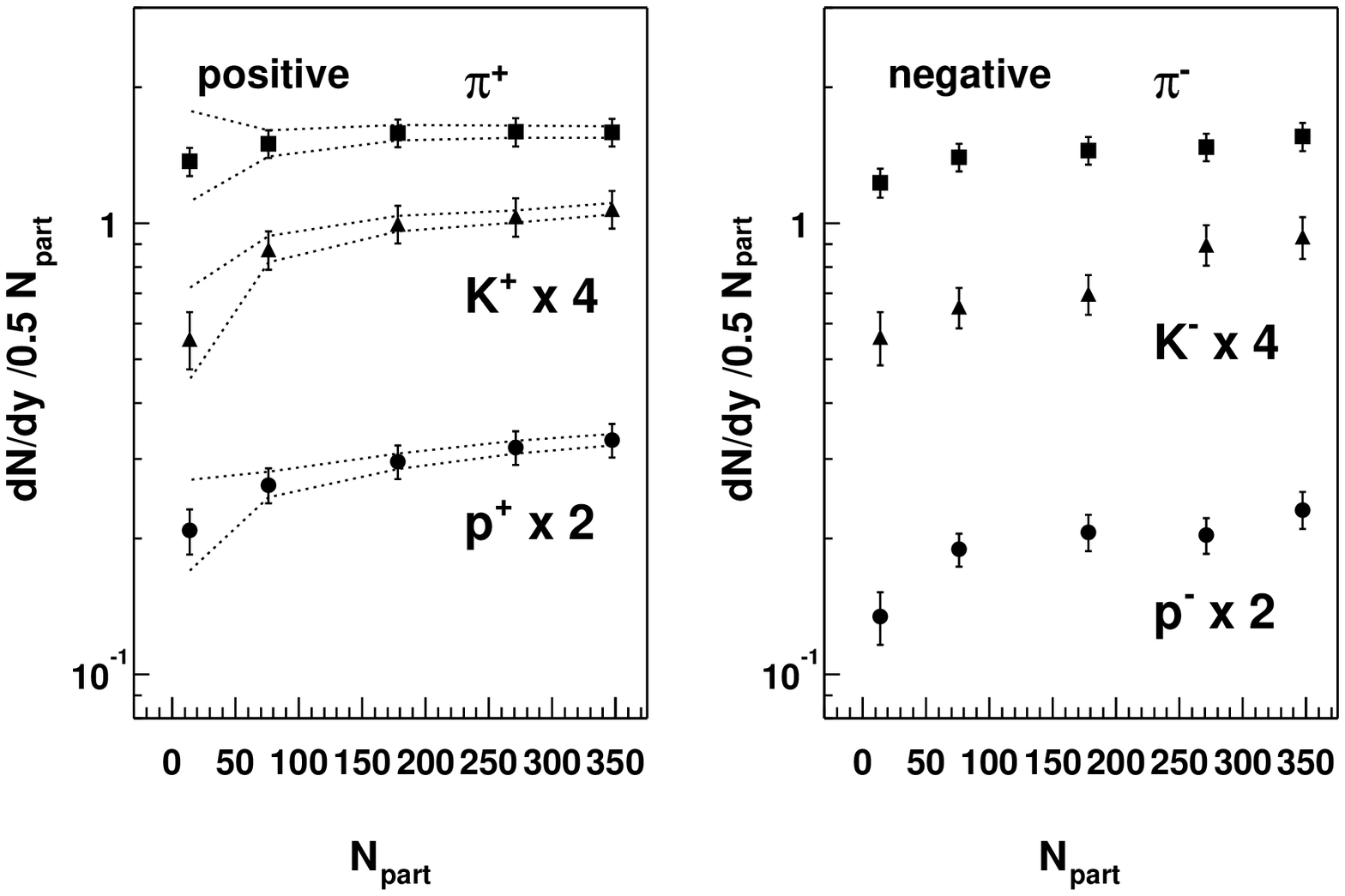}
}
\vspace*{-6pt}
\caption{The integrated $dN/dy$ per participant pair for $\pi^{+}$,
  K$^{+}$, p (left) and $\pi^{-}$, K$^{-}$, and $\overline{\rm p}$
  \protect\cite{PPG006}.  The dashed lines (left) are the systematic
  uncertainties in the Glauber model calculation of $N_{\rm part}$.}
\label{dndy}
\centerline{
\includegraphics[width=11cm,clip]{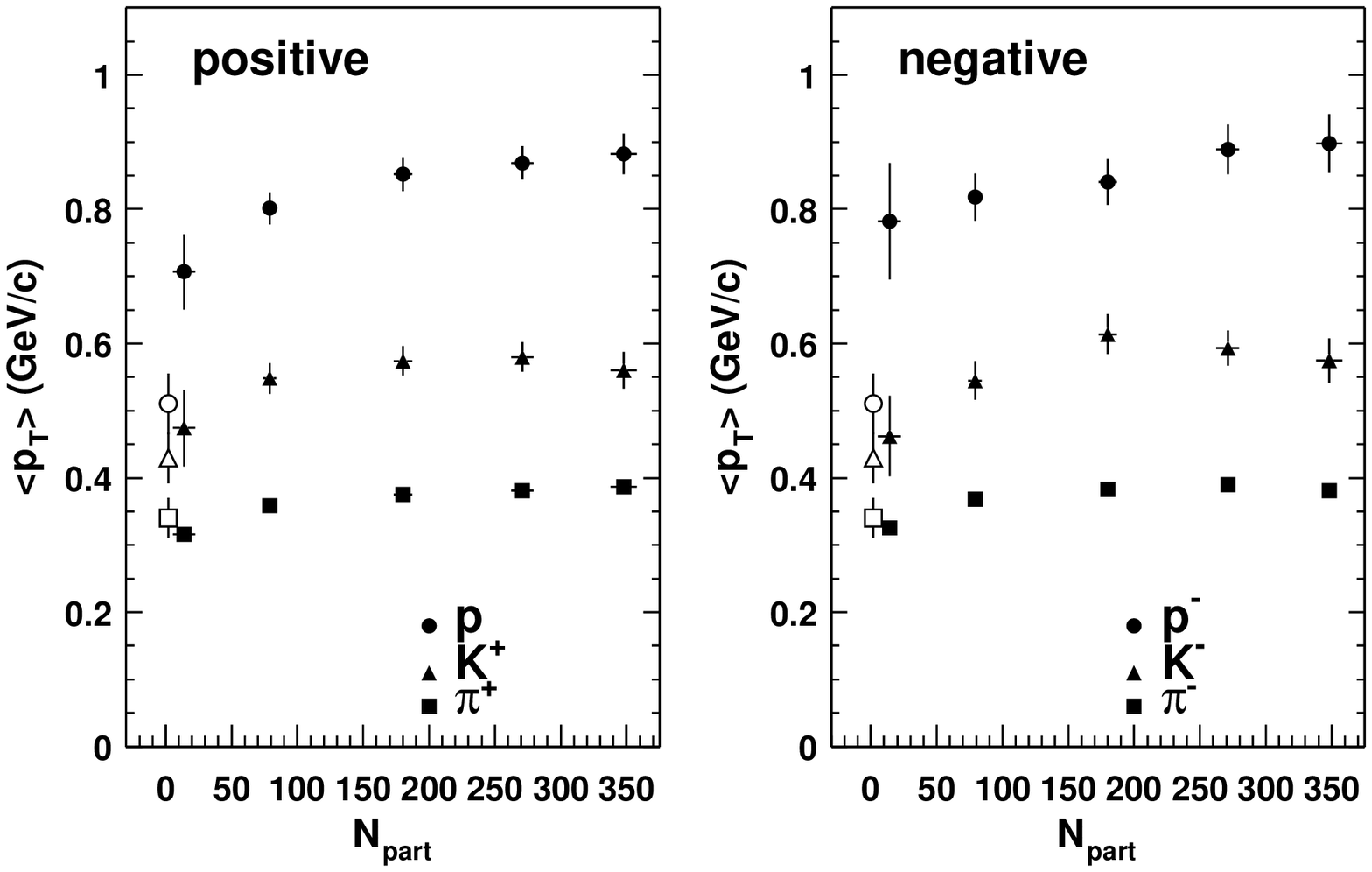}
}
\vspace*{-6pt}
\caption{
  The integrated mean ${p_{T}}$ for pions, kaons, and (anti)protons
  produced in the five different classes of event centrality
  \protect\cite{PPG006}.  The error bars are statistical only.  The
  systematic uncertainties are 7\%, 10\% and 8\% for pions, kaons
  and (anti)protons, respectively.  The open points are equivalent
  average transverse momenta from p--p and p--$\overline{\rm p}$
data, interpolated to RHIC energies.}
\label{fig3}
\vspace*{-28pt}
\end{figure}

The $dN/dy$ is normalized to pairs of participant nucleons for each
centrality selection, as shown in Fig.\ \ref{dndy}.  The dashed lines
are the systematic uncertainties in the Glauber model calculation of
the number of participant nucleons.  The kaons and (anti)protons are
scaled for clarity.  The error bars include both statistical and
systematic errors. 

The mean transverse momentum increases with the number of participant
nucleons by $20 \pm 5$\% for pions and protons, as shown in Fig.\
\ref{fig3}.  The $\langle p_{T} \rangle$ of particles produced in
p--p and p--$\overline{\rm p}$ collisions, interpolated to
RHIC energies, are consistent with the most peripheral pion and kaon
data; however, the $\langle {p_{T}} \rangle$ of protons produced in
Au+Au collisions is significantly higher.  This dependence on the
number of participant nucleons may be due to radial expansion.

\subsection{Charged particle multiplicity}

\begin{figure}[t] 
\vspace*{6pt}
\centerline{
\includegraphics[width=11cm,clip]{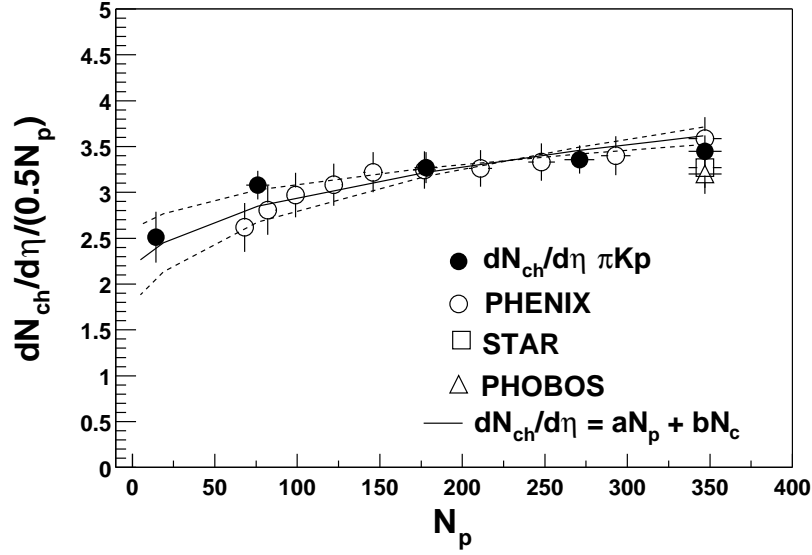}
}
\vspace*{-6pt}
\caption{The total charged multiplicity (open) in Refs
  \protect\cite{pubDndeta,star_dndeta,phobos_dndeta} and the total
  identified charged multiplicity (closed) scaled by the number of
  participant pairs are plotted as a function of the number of
  participants.}
\label{compare}
\end{figure}

The measured ${dN/dy}$ for each hadron species is converted to
${dN/d}\eta$, and the total ${dN/d}\eta$ is calculated after
summation.  In Fig.\ \ref{compare}, the resulting ${dN/d}\eta$ per
participant nucleon pair is within 5\% when compared to the
measurement made by PHENIX using the pad chambers alone \cite{PPG001}.
The results from other RHIC experiments are also shown as a
comparison.

\section{Hydrodynamic Fits}\label{hydrofits}

\subsection{Parameterization}

Radial flow would impart a velocity boost to the hadrons.  Heavy
particles should be boosted to higher ${p_{T}}$, depleting the cross
section at lower ${p_{T}}$, and resulting in a higher inverse slope in
the spectrum.  In order to measure the amount of radial expansion, a
hydrodynamics inspired parameterization is used to fit all spectra
simultaneously.  The functional form used describes a boosted thermal
source, and is based on relativistic hydrodynamics, see U. Heinz et
al.\ in \cite{sollfrank}. The particles are assumed to collectively
expand with a velocity profile that is linear with the radial
position.  The particle density distribution is assumed to be a
Gaussian in the radial position.  In addition to longitudinal boost
invariance, it is also assumed that all particles decouple
kinematically on the freezeout hypersurface at the same freezeout
temperature $T_{\rm fo}$.  The transverse velocity profile is
parameterized as:
\begin{equation}
\beta_{{T}}(\xi) = \beta_{{s}}\xi^{{n}} \,, 
\label{prof}
\end{equation}
where $\xi = {{r}}/{{R}}$, $\beta_{{s}}$ is the surface velocity, $n$ is
the profile, and $R$ is the radius of the expanding source at freezeout
($0<\xi<1$ in the case of a flat particle density distribution)
\cite{esumi_55}.  Each fluid element is locally thermalized and is
given a transverse rapidity boost $\rho$ that depends on the radial
position as:
\begin{equation}
\rho = \xi^{{n}} \tanh^{-1} \left( \beta_{{s}}\right)  \, .
\label{rho}
\end{equation}
The $m_T$ dependence of the yield ${{dN}}/{{m_T}{dm_T}}$ is
calculated after integrating over the azimuthal, rapidity, and a final
numerical integration over the radial components of the source
\begin{equation}
\frac{{dN}}{{m_{T} dm_{T}}} = \\
{A} \int{ {m_{T}} {f}(\xi) {K_1} \left( \frac{ {m_{T}} 
\cosh(\rho) }{ {T_{\rm fo}} } \right) {I_0} 
\left( \frac{{p_T} \sinh(\rho)}{{T_{\rm fo}}} \right) \xi
        {d}\xi}\,, 
\label{hydro}
\end{equation}
using the modified Bessel functions ${I_{0}}$ and ${K_{1}}$.  The
parameters in Eq.\ (\ref{hydro}) are the freezeout temperature
${T_{\rm fo}}$, the normalization $A$ and the maximum surface
velocity $\beta_{{s}}$.  A linear velocity profile (${n}=1$) within
the source is used. This choice was guided by the profile observed in a
full hydrodynamical calculation by Kolb and Heinz.$^b$ 

\subsection{Simultaneous fits to data}

\begin{figure}[!h] 
\vspace{6pt}
\centerline{
\includegraphics[width=12.5cm,clip]{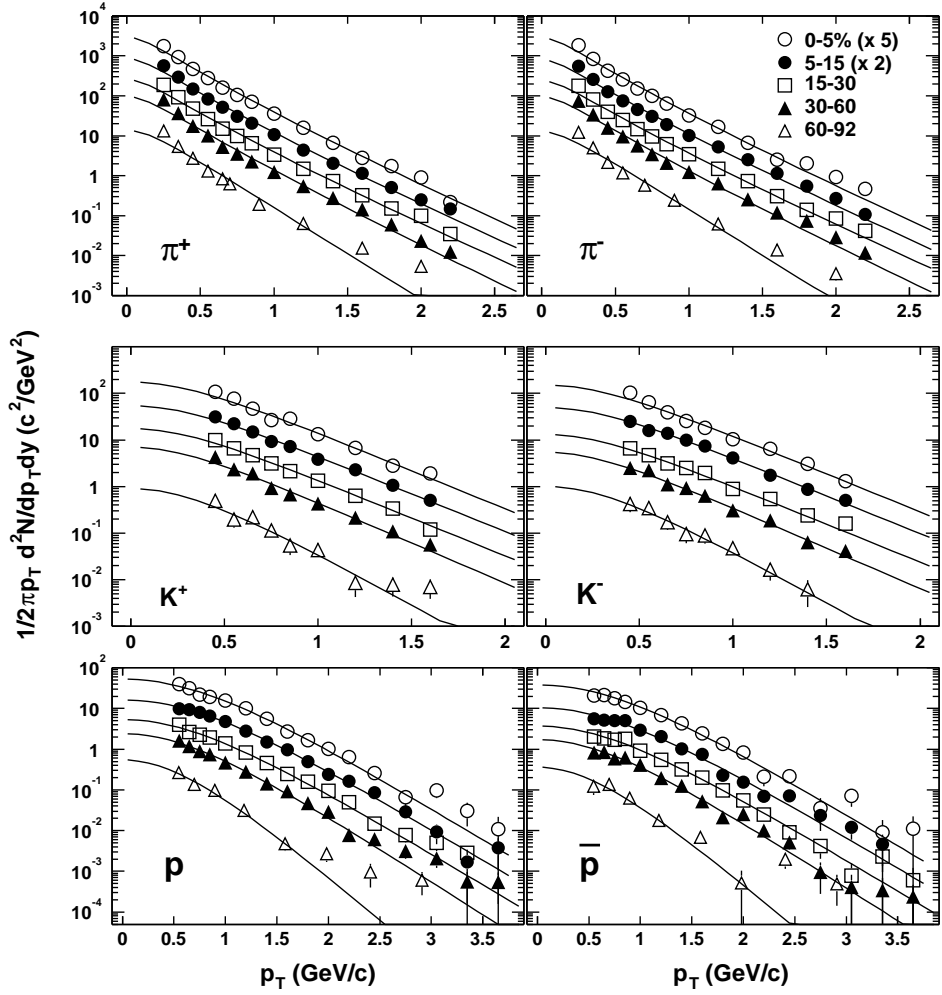}
}
\vspace*{-6pt}
\caption{The hydro-inspired parameterization is fit simultaneously 
  to the data in each event centrality}
\label{fits}
\end{figure}

The amount of radial expansion is determined by fitting all particle
transverse momentum spectra simultaneously using Eq.\ (\ref{hydro})
for $p_{T}<1.0$~GeV/c.  We make a grid of combinations of temperature
and velocity, and perform a chi-squared minimization to extract the
normalization, $A$, for each particle type. The fit is done
simultaneously for all particles, with $\chi^{2}$ contours that show
the anti-correlation of temperature and velocity.  The resonance
region in the pion spectra is excluded in the fit,$^b$ as was done in
S+S collisions at the CERN SPS \cite{reso}.  The resulting fits are
plotted with the spectra for all centralities in Fig.\ \ref{fits}.

We find that the transverse momentum distributions are consistent with
an expanding system with a surface velocity of $\beta_{{s}} = 0.72 \pm
0.01$ which decouple at a common temperature of ${T_{\rm fo}} = 122
\pm 4$ MeV.  For the 5--15\% centrality, the fit results are ${T_{\rm
    fo}} = 125 \pm 2$ MeV and $\beta_{{s}} = 0.69 \pm 0.01$.

In \cite{PPG003}, the hadron multiplicities at high ${p_{T}}$ are
different for mesons and baryons.  The suppression of pion yields at
high ${p_{T}}$ and the boost of low ${p_{T}}$ protons to high
${p_{T}}$ from radial flow may explain the dominant proton and
antiproton yields at ${p_{T}} > 1.5$ GeV/c and $1.8$ GeV/c,
respectively.

\subsection{Systematic uncertainties}

We find that Eq.\ (\ref{hydro}) is consistent with a full hydrodynamic
calculation when a linear velocity profile is used \cite{PPG009}.
Nevertheless, a parabolic profile results in an increase of
$\approx$13\% in $\beta_{{s}}$ and $\approx$5\% in ${T_{\rm fo}}$.  A
constant particle density distribution used with a linear velocity
profile decreases $\beta_{{s}}$ by $\approx$2\% with a negligible
difference in the temperature ${T_{\rm fo}}$.  As a test of the
assumption that all the particles freezeout at a common temperature,
the simultaneous fits were repeated for pions and protons only.  The
difference in ${T_{\rm fo}}$ is negligible within the measured
uncertainties.  The effect of including the pion resonance region in
the simultaneous fit decreases ${T_{\rm fo}}$ by 20 MeV.

\section{Conclusions}\label{concl}

Identified charged hadrons produced in Au+Au collisions at 130 GeV in
five different centrality selections have been measured.  Proton and
antiproton yields at high ${p_{T}}$ dominate over pion yields in
central collisions when compared to the PHENIX measured $\pi^{0}$ data
up to ${p_{T}} = 4$ GeV/c.  A combination of suppression of pion
yields at high ${p_{T}}$ and the boost of low ${p_{T}}$ protons to
high ${p_{T}}$ from radial flow may be an explanation.
 
\section*{Acknowledgements}

This work was performed under the auspices of the U.S. Department of
Energy by the University of California, Lawrence Livermore National
Laboratory under Contract No.\break W-7405-Eng-48.

\section*{Notes}
\begin{notes}
\item[a] E-mail: janebh@llnl.gov
\item[b] The resonance region affects the spectral shape at
  ${p_{T}}<0.5$ GeV/c based on model calculations by Peter Kolb and
  Ulrich Heinz (private communication).
\end{notes}

\vfill\pagebreak
\pagestyle{empty}
\phantom .
\vfill\eject

\begin{thebibliography}{99}  

\bibitem{QGP} E.V. Shuryak, {\it Phys. Rep.} {\bf 61} (1980) 71.

\bibitem{BB} J.D. Bjorken, {\it Phys. Rev. D} {\bf 27}, No.\ 1 (1983).
  
\bibitem{EOS} C.M. Hung and E. Shuryak, {\it Phys. Rev. C} {\bf 57}
  (1998) 1891.
  
\bibitem{NA44_78} I. Bearden et al., {\it Phys. Rev. Lett.} {\bf 78}
  (1997) 2080.
  
\bibitem{NA49} A. Appelshauser et al., {\it Eur. Phys. J. C} {\bf 2}
  (1998) 661.

\bibitem{hijing} X.N. Wang, {\it Phys. Rep.} {\bf 280} (1997) 287.
  
\bibitem{PPG003} K. Adcox et al., {\it Phys. Rev. Lett.} {\bf 88}
  (2002) 022301.
  
\bibitem{ref1}D. Morrison et al., {\it Nucl. Phys. A} {\bf 638} (1998)
  565.
  
\bibitem{ref2}W. A. Zajc, {\it Nucl. Phys. A} {\bf 698} (2002) 39.
  
\bibitem{ref3}K. Adcox et al., to be submitted to {\it Nucl. Instr.
    Meth. A} (2002).
  
\bibitem{PPG001}K. Adcox et al., {\it Phys. Rev. Lett.} {\bf 86}
  (2001) 3500.
  
\bibitem{GLAUBER} K. Adcox et al., {\it Phys. Rev. Lett.} {\bf 88}
  (2002) 022301.
  
\bibitem{PPG006} K. Adcox et al., submitted to {\it Phys. Rev. Lett.}
  (2001) [nucl-ex/0112006].
  
\bibitem{PPG009} K. Adcox et al., to be submitted to {\it Phys. Rev.
    C} (2002).
  
\bibitem{sollfrank} E. Schnedermann, J. Sollfrank and U. Heinz, {\it
    Phys. Rev. C} {\bf 48} (1993) 2462.
  
\bibitem{esumi_55} S. Esumi, S. Chapman, H. van Hecke and N. Xu, {\it
    Phys. Rev. C} {\bf 55} (1997) R2163.
  
\bibitem{pubDndeta} K. Adcox et al., {\it Phys. Rev. Lett.} {\bf 86}
  (2001) 3500.
  
\bibitem{star_dndeta} C. Adler et al., {\it Phys. Rev. Lett.} {\bf
    87} (2001) 112303.
  
\bibitem{phobos_dndeta} B.B. Back et al., {\it Phys. Rev. Lett.} {\bf
    85} (2000) 3100.
  
\bibitem{reso} J. Sollfrank, P. Koch and U. Heinz, {\it Phys. Lett. B}
  {\bf 252} (1990) 256.
  
\bibitem{JEFF} J.T. Mitchell et al., {\it Nucl. Inst. Meth. A} {\bf
    482} (2002) 498.

\end{thebibliography}
\end{document}